\documentstyle[aps,twocolumn]{revtex}
\author{Mario Nicodemi$^{a,b}$, Antonio Coniglio$^{a,b}$, 
Hans J. Herrmann$^{a,c}$ }
\address{
\vspace{0.2cm}
$a)$ P.M.M.H. E.S.P.C.I, 10 rue Vauquelin, 75231 Paris Cedex 05, France\\
$b)$ Dipartimento di Scienze Fisiche, Universit\'a di Napoli
  ``Federico II'', INFM and INFN Sezione di Napoli \\
Mostra d'Oltremare, Pad. 19, 80125, Napoli, Italy \\
$c)$ ICA 1, Universit\"at Stuttgart, Pfaffenwaldring 27,
70569 Stuttgart, Germany}

\title{ A model for the compaction of granular media }

\date{\today}

\newcommand{\lan}{\langle}
\newcommand{\ran}{\rangle}

\begin{document}

\maketitle

\begin{abstract}
We introduce a lattice model, in which frustration plays 
a crucial role, to describe relaxation properties of 
granular media. We show Monte Carlo results for compaction 
in the presence of vibrations and gravity, which compare well with 
experimental data.
\end{abstract}
\bigskip

Despite their importance for industrial applications
relaxation phenomena in non-thermal disordered
systems as granular media, have just recently begun 
to be studied systematically.
A common and simple experiment in this context, is the 
compaction of sand. When a box filled with loose
packed sand is shaken at low amplitude, density visibly increases. 
If in addition the density goes beyond a definite threshold the mechanical
properties of sand abruptly change and the granular
structure cannot be sheared any longer without a volume increase. 
This phenomenon, very important in practical applications
~\cite{soilmech}, was observed by
Reynolds ~\cite{Reynolds} and is referred to as the ``Reynolds''
or ``dilatancy'' transition. 

For a given macroscopic
parameter as density a granular packing can be in a huge number of 
different microscopical states. In order to describe this situation
concepts from statistical mechanics have been introduced
\cite{JNScience92,Edwards,Hans}. Relations 
to spin glasses (SG) have been suggested
several years ago (see references in~\cite{JNScience92}). 
In fact a characteristic of SG is their non trivial phase space
which gives rise to its complex static and dynamic 
behavior. The phase space structure of SG is due to the presence of 
quenched disorder and frustration. Strictly speaking 
quenched disorder is not present in granular media but
there are effects of 
``geometric frustration'', known also from hard sphere systems. This kind of
frustration is generated by the steric constraints imposed 
by the hard core repulsion of neighboring grains
and the subsequent interlocking which leads to non local cooperative 
macroscopic rearrangements. 
Recently the analogies between an intrinsically frustrated system like
frustrated percolation \cite{Coniglio} 
and phase transitions in a granular packing have 
been outlined~\cite{ConiglioHerrmann}.

In this paper we present computer simulations of a simple 
frustrated Ising lattice gas model, subject to gravity  
following a diffusion like
Monte Carlo dynamics. The particles in this system are characterized by
internal degrees of freedom which describe their orientation or other
local sterical properties of the grains. This model without gravity shows
complex behavior similar to the one observed in glass forming liquids 
and spin glasses~\cite{ConiglioNicodemi}. 
We will show how the density of our lattice gas is
strongly dependent on the duration and the amplitude of simply implemented
vibrations. Our data reproduce the logarithmic relaxation behavior
found in real experiments in a sequence of taps and offer the
possibility to make new predictions also for single tap processes. 
Our data also reproduce the distribution of 
forces at the bottom of the system as found in real experiments.
A relation appears between the SG transition,
signaled by the vanishing of macroscopic self-diffusion,
and the Reynolds transition in
granular systems. 

We consider a system of particles which move on a square lattice 
whose bonds are characterized by quenched random numbers $\epsilon_{ij}=\pm 1$. 
On site $i$ we set $n_i=1$ if a particle is present and 0 otherwise.
The particles have an internal degree of freedom $S_i=\pm 1$ and are
subjected to the constraint that whenever two ($i$
and $j$) are neighboring, their ``spin'' must satisfy the relation 
\begin{equation}
\epsilon_{ij}S_i S_j=1
\label{F}
\end{equation}
i.e. they have to fit the local ``geometrical'' 
structure. When the density of particles is high enough 
they can feel the frustration that has been imposed by the choice
of the $\epsilon_{ij}$. As a consequence, in 
resemblance to frustrated percolation~\cite{Coniglio}, 
particles can never close a frustrated loop in the lattice
leaving empty sites (see below).

The physical origin of the bond variables $\epsilon_{ij}$,
is the geometrical frustration originated in granular systems
by the actual shapes and arrangements of particles
and the internal variables $S_i$ mimic local shapes and positions. 

We have studied this system when subject to ``gravity'' and
``external vibrations''. 
The dynamics of our model consists in 
a random diffusion of particles on a square
lattice tilted by $45^{\circ}$ (see Fig.\ref{lattice}) 
in such a way as to preserve the above constraint. 
The particles attempt a move  upward with probability $P_2$ and downward with
$P_1$ (with $P_1+P_2=1$). The move is made only if the internal degrees
of freedom satisfy eq.~(\ref{F}). Similarly a spin
flips with probability one if there is no violation of eq.~(\ref{F}),
and does not flip otherwise. In absence of 
vibrations, the effect of gravity imposes
$P_2=0$. When vibrations are switched on $P_2$ becomes finite.
The crucial parameter which controls the dynamics and the final
density is the
ratio $x(t)=P_2(t)/P_1(t)$ which describes the 
amplitude of the vibration.

It is possible to associate to this model a standard Hamiltonian
formalism and establish a magnetic analogy, based on the
following definition:
\begin{equation}
-H=\sum_{\lan ij\ran } J(\epsilon_{ij} S_i S_j -1)n_i n_j +\mu \sum_i n_i
\label{H}
\end{equation}
where $S_i=\pm 1$ are spin variables, $n_i=0,1$ occupancy variables 
and $\epsilon_{ij}=\pm 1$ quenched interactions associated to the
bonds of the lattice.
It has been shown in mean field approximation~\cite{ANS} and
numerically for finite dimensional systems~\cite{ConiglioNicodemi}, 
that Hamiltonian (\ref{H}) exhibits a spin glass transition at high
density (or low temperature). 

This Hamiltonian reduces 
in the $\mu\rightarrow\infty$ limit (all sites occupied) 
to the usual {\em $\pm J$ Ising spin glass}~\cite{BY}. In the limit 
$J\rightarrow\infty$, it describes a lattice gas in which
frustrated loops entirely filled with sites are forbidden 
because along closed paths energetic reasons impose
the quantity $\sum_{i,j\in ~ loop}(\epsilon_{ij} S_i S_j -1)$ to be zero. 
In this limit a version of {\em site frustrated percolation} is recovered 
\cite{Coniglio,ConiglioNicodemi}. 
When the particle number is fixed the configuration space of the 
system obtained in this last limit is the same as that of
the frustrated lattice gas introduced at the beginning of this
paragraph. 

We have studied the model introduced above in a 2D box with periodic
boundary conditions along the x-axis and rigid walls at its bottom and 
top. 

After fixing the random quenched $\epsilon_{ij}$ on the bonds, 
a random initial particle configuration is
prepared by randomly inserting particles of given spin 
into the box from its top and then letting them fall down, with the
described dynamics ($P_2=0$), until the box is filled. 
To obtain an initial low density configuration
we do not allow particle spins to flip in
this preparation process. 
The state prepared in this way has a density of about $0.520$ and corresponds 
to a random loose packing.

We know experimentally that sand which is randomly
poured into a box reaches higher density after shaking. 
In some experiments, the shaking process occurs in a sequence of ``taps''. 
A tap is defined by its duration and its amplitude. 
After a sequence of taps the density decays logarithmically 
to a static limit (see ~\cite{Knight}). 
We have studied the phenomena of density relaxation during a
sequence of taps. In our MC simulation, 
each tap is a process in which vibrations are step like:
$x(t)=x_0$ if $t\in[0,\tau]$ while 
for $t>\tau$ the system evolves subject only to gravity 
($P_1=1$ and $P_2=0$, i.e. $x(t)=0$) until it reaches
a final ``static''configuration~\cite{fo}.  
After each tap we have measured the bulk density of the system 
$\rho(\tau,x_o;t_n)$ 
defined as the mean density in the lower 
$25\%$ of the box 
($t_n$ is the n-th tap number). Our results for
density relaxation, in a box of size $30 \times 60$
averaged over 32 different $\{ \epsilon_{ij} \}$ configurations, 
are shown in the insert of Fig.~\ref{den_tap}. The behavior of 
$\rho(\tau,x_o;t_n)$ is well fitted by the following
logarithmic function in agreement with the experimental data 
(see \cite{Knight,Ben-Naim}):
\begin{equation}
\rho(\tau,x_o;t)=\rho_{s}-\Delta\rho_{\infty}/[1+B ~ ln(t/\tau_0+1)]
\label{logrel}
\end{equation}
In Fig.~\ref{den_tap} we have collapsed our results 
for four different amplitudes as well as the
experimental data for three different amplitudes on a single
curve using eq.~(\ref{logrel})
and see that the agreement is very satisfactory.

We have also simulated a single tapping process. 
In this case we have found that the relaxation, instead of being 
logarithmic as in the sequence of taps is 
well described by a ``stretched exponential''.
In the simulation we start at $t=0$ from a random loose packing configuration
described before, then we introduce ``vibrations'' in the interval 
$t\in[0,\tau]$ linearly decreasing
the ratio $x(t)$, $x(t)=x_0(1-t/\tau)$ (with $x_0=1$).  
For $t>\tau$ we put $x(t)=0$ and let the system
evolve until it reaches a final ``static''configuration.
The final ``static'' bulk density $\rho(\tau)$ monotonically increases
with the vibration time $\tau$ asymptotically reaching a maximal 
density value $\rho^*\sim 0.78$ when $\tau\rightarrow\infty$
~\cite{our_fut}. 

During the dynamical process described above, we have recorded 
the time dependence of the mean bulk density 
$\rho(t,\tau)$. We find that the static limit is reached with 
a stretched ``relaxation form'' ~\cite{our_fut}: 

\begin{equation}
\rho(t,\tau)=\rho_s(\tau)-A~exp(-((t-t_0)/T)^\beta)
\label{strexp}
\end{equation}

Typical values of the parameters of eq.~(\ref{strexp})
in our range of
$\tau$ are $A\in [0.15,0.25]$, $t_0\in [-10^1,10^3]$, $T\in [10^2,10^4]$,
$\beta\in [2,4]$. 
Note that the stretched exponential behavior only sets in after a time
$t_0$, which can be very long if $\tau$ is long.
The relaxation processes found here are 
rather different from the logarithmic 
relaxation found in the sequence of taps and could be 
investigated experimentally.   

The effect of compaction 
is clearly shown by the final density profile as a function of depth
$h$, $\rho(h,\tau)$, depicted in Fig.\ref{denpro}. In this case the box has a
size $100 \times 200$
and the final states have been averaged over 32 to 512 different 
$\{ \epsilon_{ij} \}$
configurations (according the value of $\tau$). 
As suggested in ref.~\cite{Haya-hong} the density profile
of granular media can be fitted using a generalized
Fermi-Dirac distribution. 
As shown in Fig.\ref{denpro} the data from our model are well fitted 
by such a function for  different values of $\tau$:

\begin{equation}
\rho(h,\tau)=\rho_s(\tau)[1-1/[1+exp((h-h_0(\tau))/s(\tau))]]\ \ .
\label{fermi-dirac}
\end{equation}

To characterize a particle packing and its capability of internal
rearrangement, we  
studied their self-diffusivity at fixed global particle density 
by setting 
$x=1$. Specifically we have studied the time dependence
of the particle mean square 
displacement $R^2(t)=\lan \frac{1}{N}\sum_i (r_i(t)-r_i(0))^2 \ran$ 

A very interesting phenomenon is observed for densities close to the maximal 
value $\rho^*$: $R^2(t)$ shows 
deviations from the linear time dependence typical of standard
Brownian diffusive motion and presents an inflection 
point\cite{ConiglioNicodemi}. 
This signals the
existence of two characteristic time regimes for particle motion (as
already argued in \cite{Mehta-Barker-Duke}). 
From the long time behavior of $R^2(t)\sim Dt$ 
we extract the diffusion 
coefficient $D(\rho)$, which goes to zero at about 
$\rho^*$, signaling a localization transition in which particles are
confined in local cages and the macroscopic diffusion-like 
processes are suppressed. 
This phenomenon may also be described 
in a different way: $\rho^*$ is
the density above which it becomes impossible to obtain a macroscopic
rearrangement of the particle positions without increasing the system volume,
i.e. the density at which macroscopic shear in the system is
impossible without dilatancy. This then seems to correspond
to the quoted Reynolds transition in real granular media.

The density $\rho^*$  coincides
with the density at which the spin glass (SG) transition of 
Hamiltonian (\ref{H}) (for $J\rightarrow\infty$) is located. 
This implies that at $\rho^*$ 
the SG correlation length $\xi_{SG}$ diverges,
signaling the presence of collective behavior in the system. 
In SG this length can be detected by measuring the non linear susceptibility.
In granular material the spin variables represent internal degrees
of freedom and cannot be easily detected.

The coincidence of the SG transition and the suppression of 
self-diffusivity suggest the equivalence of the Reynolds
transition in
granular media, the SG transition in magnetic systems and the
``ideal'' glass transition in glass forming liquids
\cite{ConiglioHerrmann,ConiglioNicodemi}. Like in glass forming 
liquids the Reynolds transition does not show divergences in thermodynamic 
functions.

The model introduced here is suited to study also other aspects of
granular media. 
If a force is applied at the top of a granular system in a box, the local
distribution of forces $v$ at the bottom follows an exponential law
$P(v)=a\cdot exp(-c ~ v)$.
As suggested in
\cite{BouchaudCatesClaudin,JNScience95,EdwardsMounfield}, 
it is possible to introduce simplified models to
describe the physics of forces in granular systems. 
In particular, a model has been proposed  in \cite{JNScience95}  
in which forces are represented by scalars.
In our lattice model we
apply the same approximation to study the force distributions in 
static configurations of the system. We suppose that each present site
($n_i=1$) carries its own weight (equal to unity) and
transmits the force  $w_i$ acting on it to its first left and right
neighboring sites in the lower row. 
If its right (left) neighbor has a distance $l_r$
($l_l$) from site $i$, the force contribution it receives from this site
is equal to 
$w_i \cdot l_l/(l_r+l_l)$ ($w_i \cdot l_r/(l_r+l_l)$ respectively for 
the left site). We have calculated the force distribution 
$P(v)$ at the bottom of our system as shown in Fig.~\ref{Pv}. 
In agreement with the experimental data and the result of the model 
introduced  in \cite{JNScience95}, our data are well fitted
by:
\begin{equation}
P(v)=a\cdot v^b exp(-c ~ v)
\label{vexpv}
\end{equation}

As noted in Ref.
\cite{JNScience95}, 
the power law in front of the exponential
would be very difficult to be detected experimentally since it affects the 
distribution for small value of $v$.

In conclusion in this paper a frustrated Ising lattice gas has been 
introduced to describe different 
aspects of the phenomenology of granular systems,
such as compaction in the presence of vibrations, logarithmic density
relaxation, and exponential force distribution. 
The results are in agreement
with real experiments. The model is able to predict new results 
which are amenable to experimental observation, some of which have been 
reported here, while others are under investigation ~\cite{our_fut}.
The model which contains geometrical frustration as essential 
ingredient also shares features of spin glasses and glass forming
liquids.

Although we have reported here numerical results in 2D we expect the
same features also in 3D. 

\bigskip

We thank H. Jaeger and J. Knight for sending us their
experimental data and IDRIS for computer time on Cray-T3D. 


\begin{figure}[h]
\caption{Schematic picture of the lattice model considered 
  here. Wavy and straight lines represent the two different kinds
  of bonds ($\epsilon_{ij}=\pm 1$). Filled (empty) circles are present
  particles with spin $S_i=+1$ ($S_i=-1$).} 
\label{lattice}
\end{figure}

\begin{figure}[h]
\caption{
Experimental data from [11] (square) and our MC data (circle) rescaled
according eq.~(\ref{logrel}).
Insert: density $\rho(\tau,x_0;t_n)$ from our MC data 
as a function of tap number $t_n$, 
for tap vibrations of amplitude $x_0=0.001,0.01,0.05,0.1$ (from bottom to
top) and duration $\tau=3.28\cdot 10^1$. 
The superimposed curves are logarithmic fits from 
eq.~(\ref{logrel}). 
}
\label{den_tap}
\end{figure}

\begin{figure}[h]
\caption{The density profile $\rho(h,\tau)$ as a function of depth $h$
($h=0$ corresponds to the top of the box, $h=200$ to the bottom) for 
different values of the vibration duration $\tau$
($\tau\in[3.28\cdot 10^{-3},4.92\cdot 10^{4}]$). 
In the bulk of the system, for fixed $h$, 
$\rho(h,\tau)$ is an increasing function of $\tau$.
Continuous lines are
Fermi-Dirac function fits from eq.~(\ref{fermi-dirac}).
Insert: Rescaled density profile $\rho(h,\tau)/\rho_s(\tau)$ as a function
  of the rescaled depth $(h-h_0(\tau))/s(\tau)$ ($\rho_s(\tau)$, 
$h_0(\tau)$ and $s(\tau)$, 
are fitting parameters to obtain the data collapse).}
\label{denpro}
\end{figure}

\begin{figure}[h]
\caption{Force distribution $P(v)$ as a function of weight 
$v$ normalized by the mean force felt by the sites, for a static 
configuration of density $\rho_s=0.764$. 
Superimposed is the fit function in 
eq.~(\ref{vexpv}). The fit parameters are $a=12.4$, $b=5.6$ and $c=4.6$.
The distribution $P(v)$ becomes narrower when the bulk density 
increases and is independent of the depth at which is measured (see [18]).} 
\label{Pv}
\end{figure}


\begin{thebibliography}{40}
\bibitem{fo} A static configuration is defined by the criterion
that during a fixed time $t_{repose}$ does not change any
longer. We fixed in our simulation $t_{repose}=330$ time steps. 
Time is measured in such a way that unity corresponds to one single
average update of all particles and all spins of the lattice.
\end{thebibliography}
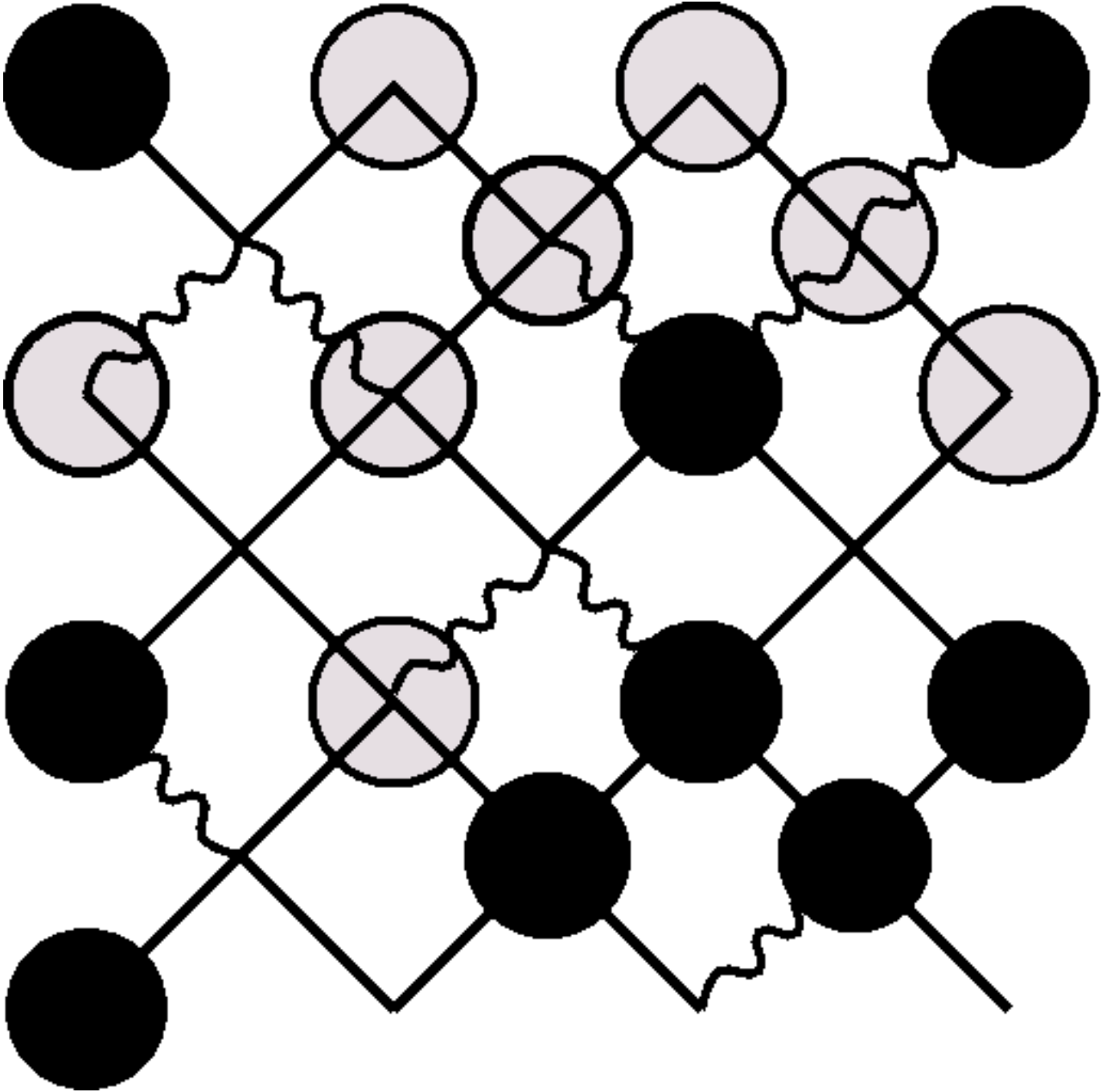
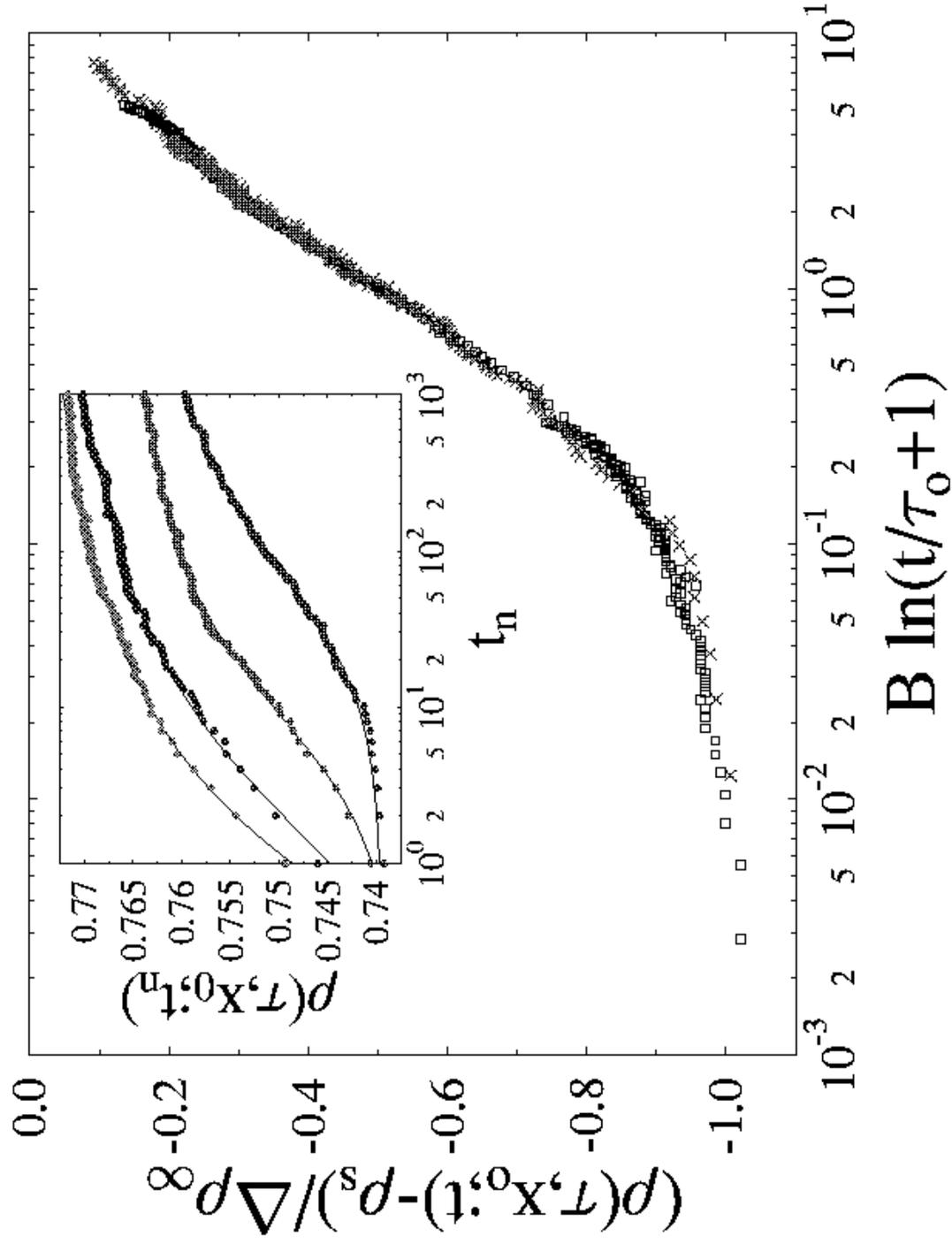
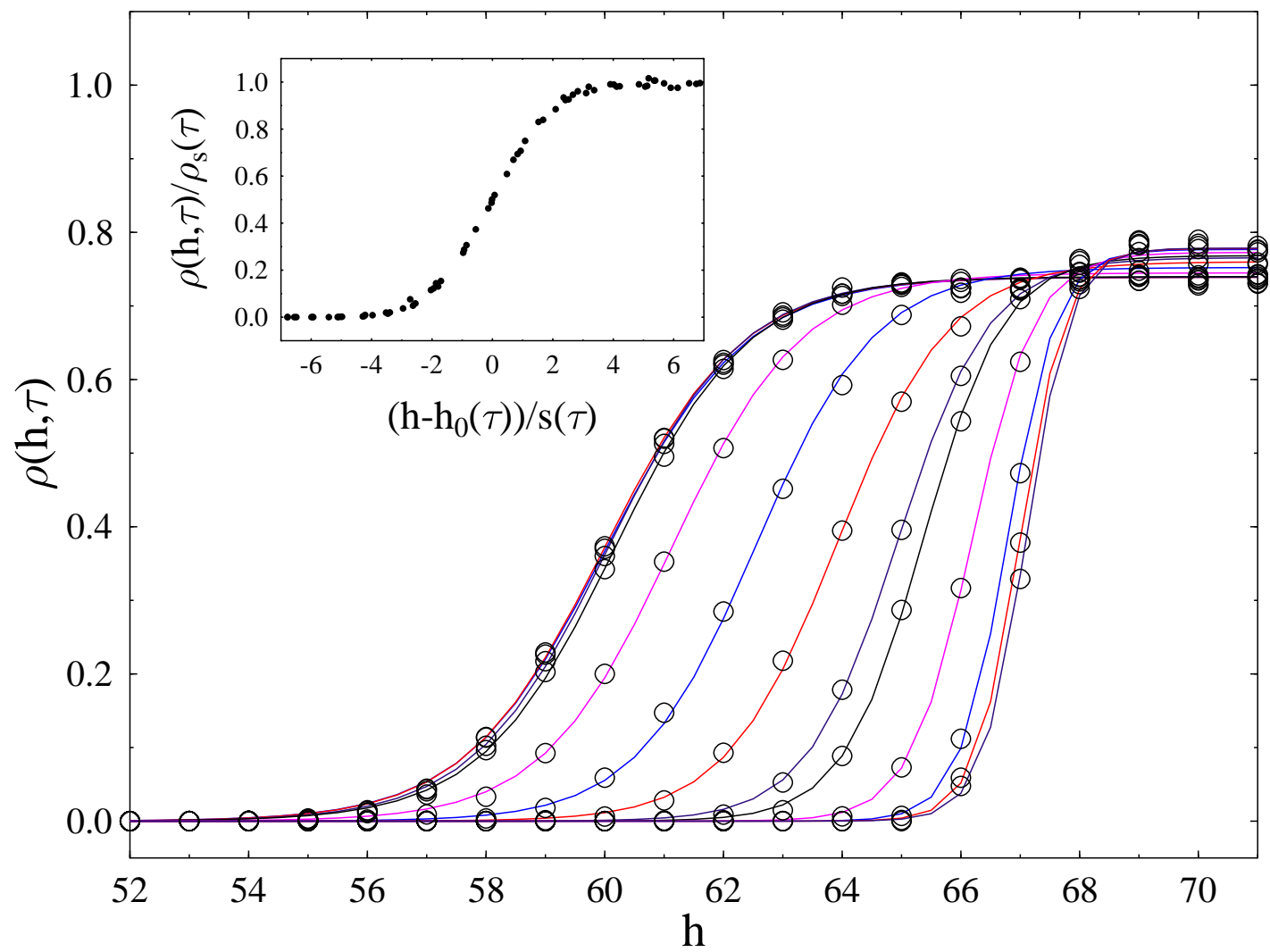
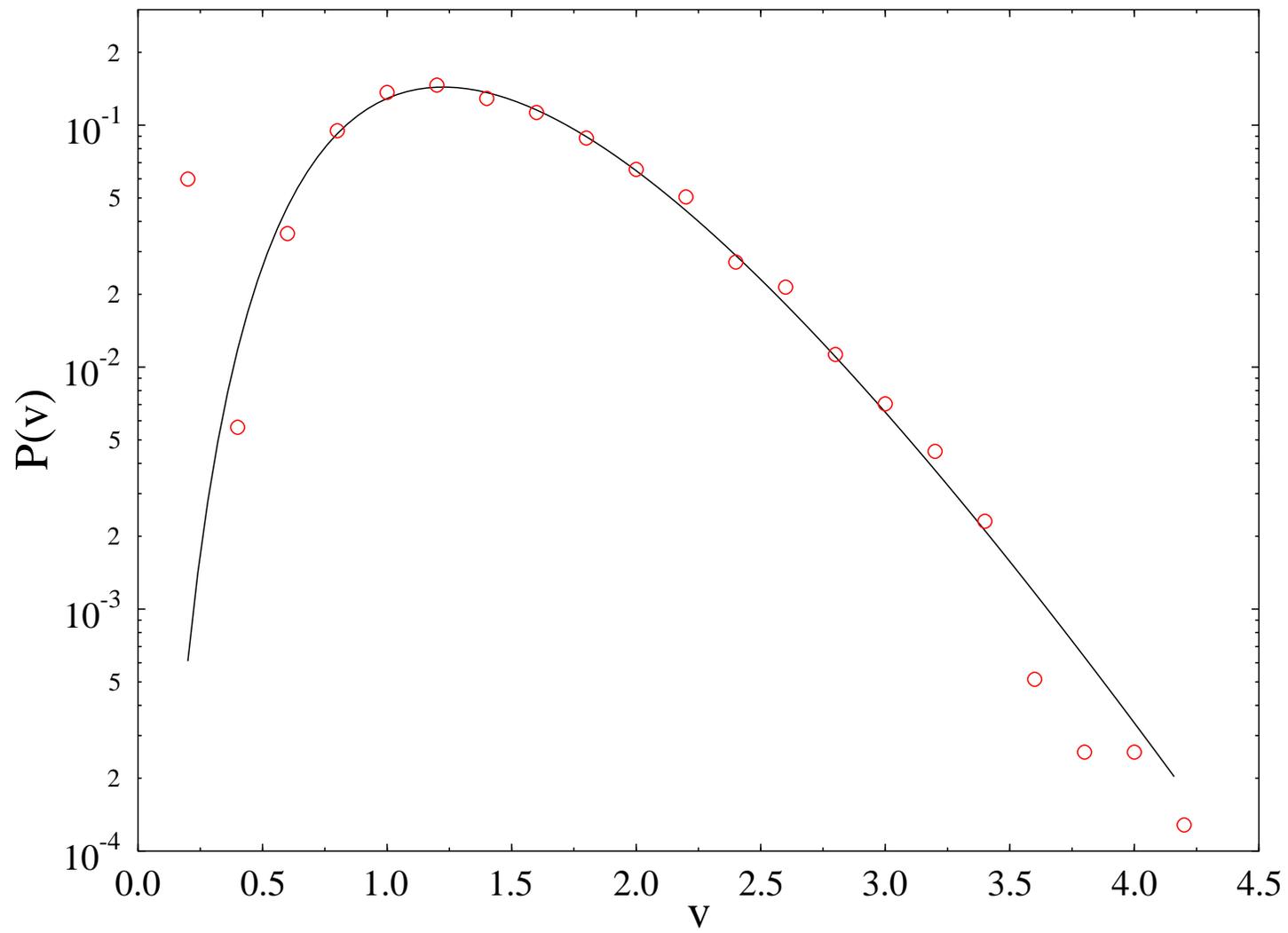
\end{document}